\documentclass{eptcs}
\providecommand{\event}{UITP'16} % Name of the event you are submitting to
\usepackage[utf8x]{inputenc}
\usepackage[T1]{fontenc}
\usepackage[american]{babel}

\usepackage[scaled=0.88]{helvet}
\usepackage{microtype}

\usepackage{hyperref}
\usepackage{xspace}
\usepackage{graphicx}
\usepackage{xcolor}
\usepackage{caption}
\usepackage{listings}

\hypersetup{
  colorlinks = true,
  citecolor=purple,
  linkcolor=blue
}

\newcommand{\aeg}{\textsf{Alt-Ergo}\xspace}
\newcommand{\agr}{\textsf{AltGr-Ergo}\xspace}

\definecolor{bg}{rgb}{0.95,0.95,0.95}

\makeatletter

\usepackage[framemethod=tikz]{mdframed}
\usepackage[listings,skins]{tcolorbox}

\def\beginlstdelim#1#2#3#4%
{%
  \def\endlstdelim{#2\egroup}%
  {\ttfamily#3#1}\bgroup#4\aftergroup\endlstdelim%
}

\lstdefinelanguage{ocaml}{
  language=[Objective]caml,
  identifierstyle=\ocidstyle
}

\lstdefinelanguage{altergo}{
  alsoletter={<}{:},
  morekeywords={type,logic,axiom,goal,let,forall,exists,in},%
  classoffset=1,
  morekeywords={int,real,bool,set,prop},keywordstyle=\color{blue!60!black}\bfseries,
  classoffset=2,
  morekeywords={Error:,Warning:},keywordstyle=\color{red}\bfseries,
  classoffset=0,
  sensitive, %
  moredelim=**[is][\beginlstdelim{axiom\ }{:}{\color{green!50!black}\bfseries}{\color{blue!80!black!50!white}\bfseries}]{axiom\ }{:},
  moredelim=**[is][\beginlstdelim{logic\ }{:}{\color{green!50!black}\bfseries}{\color{blue!80!black!50!white}\bfseries}]{logic\ }{:},
  moredelim=**[is][\beginlstdelim{goal\ }{:}{\color{green!50!black}\bfseries}{\color{blue!80!black!50!white}\bfseries}]{goal\ }{:},
  morecomment=[n]{(*}{*)},%
  morestring=[d]",%
  literate={=>}{{$\Rightarrow$}}1
  {<=>}{{$\Leftrightarrow$}}1
  {>=}{{$\geq$}}1
  {=<}{{$\leq$}}1
  {->}{{$\rightarrow$}}1
  {<-}{{$\leftarrow$}}1
  {<->}{{$\leftrightarrow$}}1
  {forall}{{\color{blue!60!black}\bfseries$\forall\,$}}1
  {exists}{{\color{blue!60!black}\bfseries$\exists\,$}}1
  {<>}{{$\neq$}}1
  {and}{{$\wedge$}}1
  {or}{{$\vee$}}1
  {not}{{$\neg$}}1
  {'a}{{\color{blue!60!black}$\alpha$}}1
  {'b}{{\color{blue!60!black}$\beta$}}1
  {'c}{{\color{blue!60!black}$\gamma$}}1
}

\newcommand*\ocidstyle{%
        \expandafter\id@style\the\lst@token\relax
}
\def\id@style#1#2\relax{%
        \ifcat#1\relax\else
                \ifnum`#1=\uccode`#1%
                        \color{blue!60!black}
                \fi
        \fi
}

\newenvironment{tcb}[2][\small]{%
  \tcblisting{%
    listing only,colback=bg,colframe=bg,enlarge
    top by=0mm,top=0pt,bottom=0pt,left=2pt,right=2pt,
    before={\par\vspace{7pt}\noindent},
    after={\par\vspace{7pt}\noindent},
    listing options={language=#2,basicstyle={\ttfamily#1\upshape}}%
    }}{\endtcblisting}

\newenvironment{tcbfl}[2][\linewidth]{%
  \tcblisting{%
    listing only,colback=bg,colframe=bg,enlarge
    top by=0mm,top=0pt,bottom=0pt,left=2pt,right=2pt,%
    width=#1,%
    listing options={language=#2}%
    }}{\endtcblisting}

\makeatother

\title{AltGr-Ergo, a Graphical User Interface for\\the SMT Solver Alt-Ergo}

\author{
Sylvain Conchon
\institute{LRI, Universit\'e Paris-Sud\\Orsay, France}
\email{conchon@lri.fr}
\and
Mohamed Iguernlala
\institute{OCamlPro SAS\\Gif-sur-Yvette, France}
\email{mohamed.iguernlala@ocamlpro.com}
\and
Alain Mebsout
\institute{The University of Iowa\\Iowa City, USA}
\email{alain-mebsout@uiowa.edu}
}

\def\titlerunning{AltGr-Ergo}
\def\authorrunning{S. Conchon, M. Iguernlala \&  A. Mebsout}

\begin{document}

\maketitle

\begin{abstract}
  Due to undecidability and complexity of first-order logic, SMT
  solvers may not terminate on some problems or require a very long
  time.  When this happens, one would like to find the reasons why the
  solver fails. To this end, we have designed \agr, an interactive
  graphical interface for the SMT solver \aeg which allows users and
  tool developers to help the solver finish some proofs. \agr gives
  real time feedback in order to evaluate and quantify progress made
  by the solver, and also offers various syntactic manipulation
  options to allow a finer grained interaction with \aeg. This paper
  describes these features and their implementation, and gives usage
  scenarios for most of them.
  \end{abstract}

\section{Introduction}

\aeg is an SMT solver designed for checking logical formulas generated
by deductive program verification frameworks.  For instance, \aeg is
used as a back-end in the \textsf{Why3} plateform~\cite{why3}. It is
also used to discharge formulas derived from C programs in
\textsf{Frama-C}~\cite{framac2015}, from Ada programs in
\textsf{SPARK}~\cite{spark} or from B machines in
\textsf{Atelier-B}~\cite{AtelierB, AE-RSSR-2016}.

The \aeg input files produced by such tools share the same
structure. They start with a prelude that contains a set of
definitions (datatypes and logical symbols) and axioms for the
encoding of theories specific to program verification (complex data
structures, memory models, etc.). The rest of the file contains a
proof obligation (PO) generated by a weakest precondition calculus.

\aeg checks such input files by the use of a combination of decision
procedures (SAT, simplex, congruence closure, etc.) for reasoning
about builtin theories (Booleans, arithmetics, equality, etc.) and a
matching algorithm for instantiating quantified formulas.

Due to undecidability of first-order logic, \aeg may not terminate on
some problems.  When this happens, one would like to find the reasons
why the solver fails. Most of the time, \aeg is either overwhelmed by
a huge number of useless instances of axioms (causing a high activity
in its decision procedures), or it fails to produce the good
instances of lemmas that are mandatory to prove a goal.

\begin{figure}[htb]
  \centering
  \includegraphics[width=1.0\textwidth]{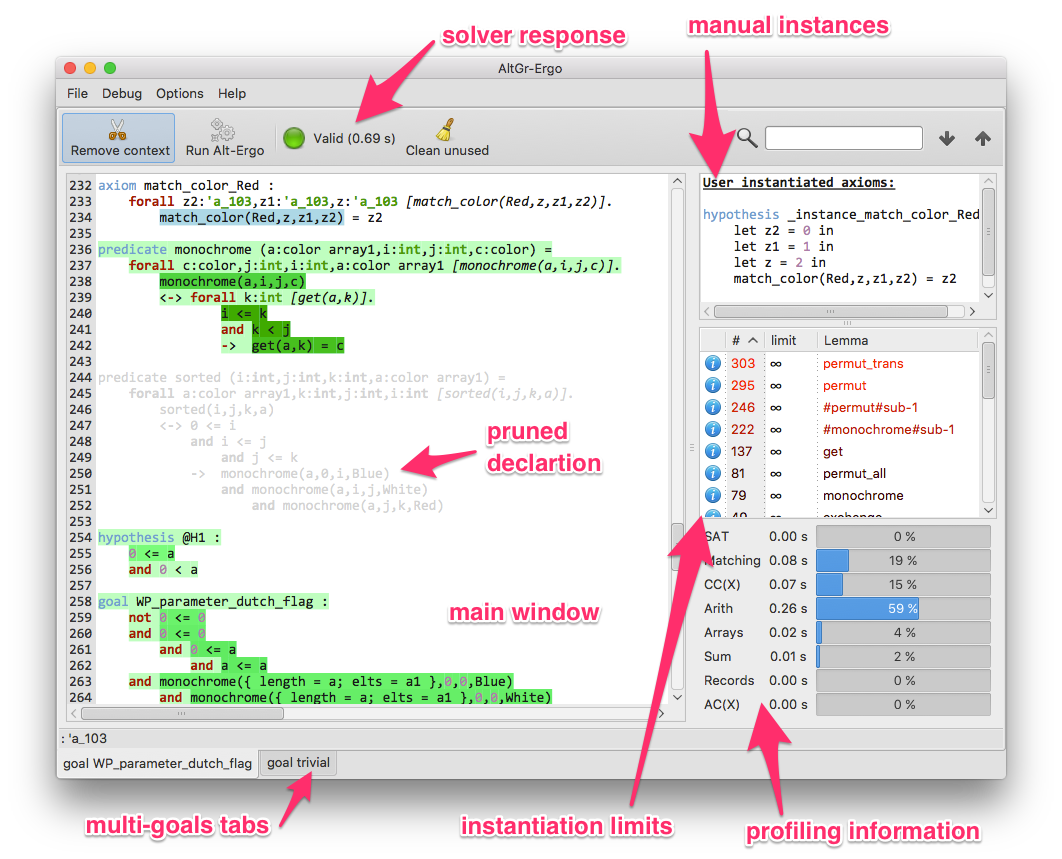}
  \caption{Overview of \agr's interface}
  \label{fig:overview}
\end{figure}

A possibility for inspecting the internals of the solver is to output debugging
information. This is however impractical because there is simply too many
things to display and the output rapidly becomes unreadable. To help users (or
developers) find out and understand what is going on, we have designed \agr, a
graphical user interface for \aeg. As shown in Figure~\ref{fig:overview}, our
GUI displays at runtime crucial profiling information about the internal
activities of the solver (time spent in decision procedures, number of
instantiations, etc.). Some interaction features have also been added so that
one can manually help the solver prove a goal (manual instances of lemmas,
selection of hypotheses, etc.).

The main features of \agr are described in the next sections.
The interface can display the following profiling information:
\begin{itemize}
\item unsat cores (Section~\ref{sec:unsat})
\item number of instances produced by axiom (Section~\ref{sec:instances})
\item time spent in decision procedures (Section~\ref{sec:profiling})
\end{itemize}
Interactive features include the following syntactic manipulations:
\begin{itemize}
\item pruning operations (Sections~\ref{sec:pruning} and
  \ref{sec:dep}) for deactivating some part of the prelude;
\item manual instances of lemmas (Section~\ref{sec:manual});
\item selection and modification of triggers
  (Section~\ref{sec:triggers}) to change the heuristics used to guide
  the matching algorithm.
\end{itemize}
Last but not least, \agr provides a session mechanism
(Section~\ref{sec:sessions}) which allows a user to save and replay
all his modifications (selections, manual instances, etc.) on a given
problem.

\section{A Short Introduction to Alt-Ergo}
\label{sec:matching}

In order to understand some aspects of the graphical interface, we
briefly present in this section \aeg{}'s syntax and a high level overview of
its main components.

The input language of \aeg{} is an extension of first-order logic with
builtin theories and prenex polymorphism\footnote{Type variables, if
  any, are prenex and implicitly universally quantified.} \emph{\`a
  la} ML~\cite{AE_polymorphism}.  Figure~\ref{fig:ex} shows a small
proof obligation written in \aeg's syntax: first, a type symbol
\texttt{set} parameterized by a type variable $\mathtt{\alpha}$ is
declared. Then, a polymorphic function (\textit{resp.} predicate)
symbol \texttt{add} (\textit{resp.}  \texttt{mem}) is
introduced. After that, an axiom \texttt{mem\_add} that gives the
meaning of membership over the \texttt{add} symbol is stated. Finally,
two integer symbols \texttt{a} and \texttt{b}, two sets of integers
symbols \texttt{s1} and \texttt{s2}, and a goal are given.

In addition to the Boolean connective ``$\to$", the ``toy goal'' mixes
symbols from two theories: the free theory of equality (\texttt{mem},
\texttt{add}, \texttt{a}, \texttt{b}, ...), and linear arithmetic
(\texttt{+, -, 1}). It is made of two parts: the hypotheses \texttt{a
  = b + 1} and \texttt{s2 = add(b,s1)}, and the conclusion
\texttt{mem(a - 1, add(b,si))} we would like to prove valid. Thanks to
the second hypothesis and a \textit{ground instance} of axiom
\texttt{mem\_add} (where {\tt x} is replaced by {\tt a - 1}, {\tt y}
by {\tt b}, {\tt s} by {\tt s1} and $\alpha$ by {\tt int}), the
conclusion is equivalent to \texttt{(a-1 = b $\lor$ mem(a-1,
  s1))}. Moreover, the latter formula always holds because \texttt{a -
  1 = b} is equivalent to the first hypothesis modulo linear
arithmetic. We thus conclude that the goal is valid.
\begin{figure}[htb]
  \centering
\begin{minipage}[h]{9cm}
\hrule
\begin{tcbfl}[9cm]{altergo}
type 'a set

logic add: 'a, 'a set -> 'a set
logic mem: 'a, 'a set -> prop

axiom mem_add:
  forall x, y: 'a. forall s: 'a set.
    mem(x, add(y, s)) <-> (x = y or mem(x, s))

logic a, b: int
logic s1, s2: int set

goal g:
  a = b + 1 ->
  s2 = add(b,s1) ->
  mem(a - 1, s2)
\end{tcbfl}
  \caption{An example problem in \aeg's syntax}
  \label{fig:ex}
\hrule
\end{minipage}
\end{figure}

% \medskip

\begin{figure}[htb]
  \centering
  \includegraphics[width=9cm]{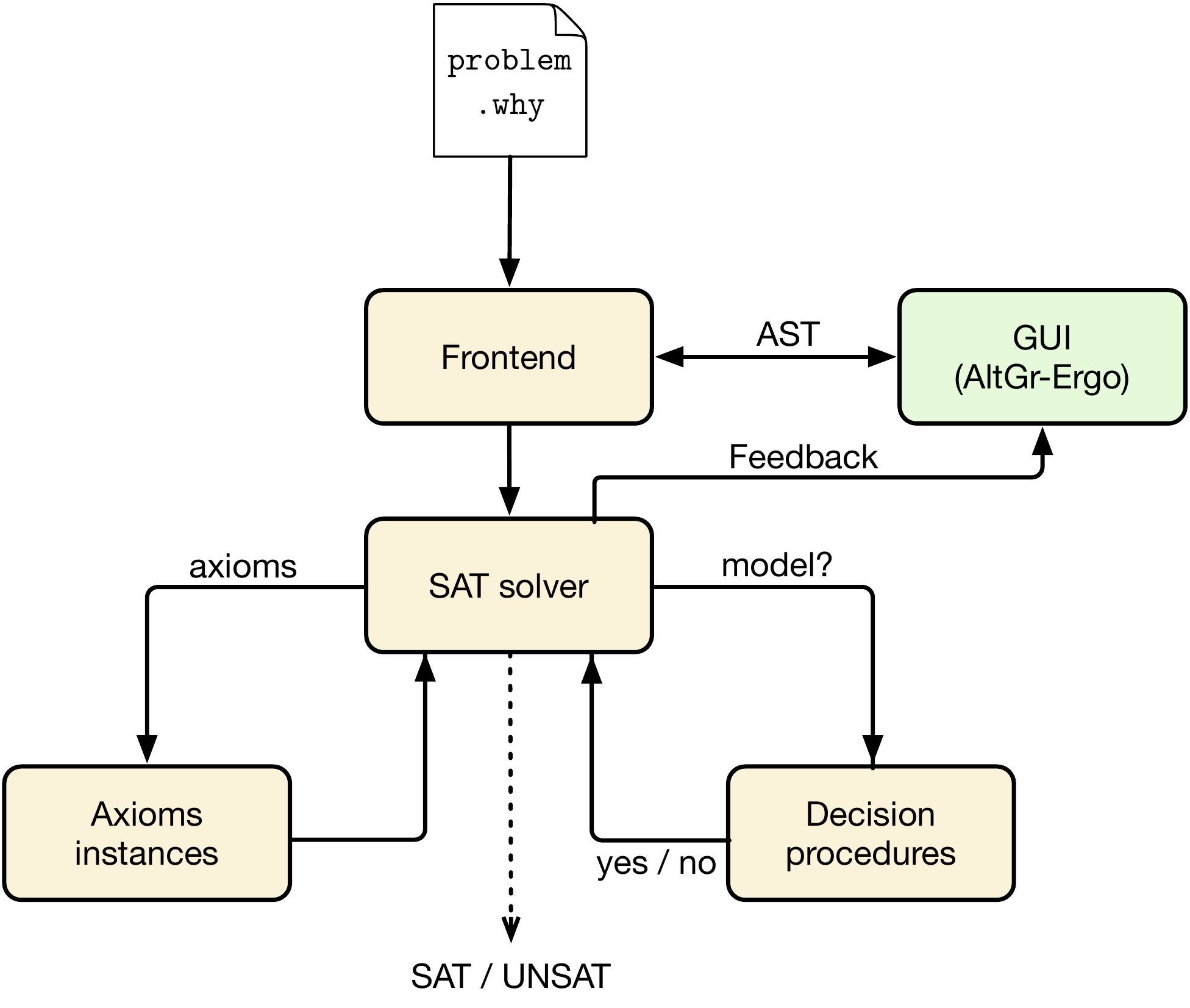}
  \caption{\aeg's simplified architecture}
  \label{fig:archi}
\end{figure}

\aeg{} handles such proof obligations following the architecture given
in Figure~\ref{fig:archi}. The solver can be called either via its
command-line ``\texttt{alt-ergo}'' or via its graphical user interface
``\texttt{altgr-ergo}''. The front end provides some basic operations
such as parsing, type-checking, triggers inference\footnote{this notion is
crucial to control how axioms are instantiated, and is
explained at the end of this section} and translation of input
formulas to richer data structures manipulated by back end modules.

The SAT solver plays a central role in \aeg{}. Given a formula, it
tries to build a (partial) Boolean model for the ground part that is
neither contradicted by the decision procedures, nor by the instances
generated from (universally quantified) axioms. Its main operations
are guessing truth values of (immediate) sub-formulas appearing in
disjunctions (\textit{decision}) and propagating unit facts that have
been deduced (\textit{bcp}). Atomic formulas (literals) are sent to
``decision procedures'' to check if they are consistent in the union
of supported theories, and universally quantified formulas are sent to
an ``axioms instantiation'' engine. If an inconsistency that does not
involve any decision is detected, the given goal is valid\footnote{To
  prove validity, \aeg{} internally assumes the negation of the
  conclusion and tries to deduce unsatisfiability.}. Otherwise, when
the SAT reaches a fix-point (\textit{i.e.} succeeds in building a Boolean
model), it asks the ``axioms instantiation'' part for some new ground
instances. These instances are added to the SAT's context and
reasoning continues.

Decision procedures component provides a combination of decision
algorithms for a collection of built-in theories. \aeg{} supports some
theories that are useful in the context of program verification, such
as the free theory of equality with uninterpreted symbols, linear
arithmetic over integers and rationals, fragments of non-linear
arithmetic, polymorphic functional arrays with extensionality,
enumerated and record datatypes, and associative and commutative (AC)
symbols. More details of our combination techniques, which are not
necessary to understand the rest of the paper, can be found
here~\cite{ccx, acx, thesisMohamed}.

To reason about axioms, \aeg{} uses an instantiation mechanism based
on E-matching~\cite{e-matching} techniques. It generates ground
consequences from assumed axioms based on some heuristics and
information provided by the SAT solver and the decision
procedures. The challenge is to heuristically produce useful instances
that will allow to discard the current SAT's model, thus reducing
the search space, and hopefully derive unsatisfiability (validity).

In addition to axioms, the instantiation engine requires a set of
ground terms, and their partition into equivalence classes (computed
by the decision procedures). In general, considered ground terms are
those that appear in the decision procedures environment when
instantiating. If this does not generate any instance, all the ground
terms that appear in the current Boolean model are considered.
Another key ingredient is the use of the notion of triggers
(\textit{a.k.a} patterns or filters) to guess which instances may be
relevant depending on the SAT and decisions procedures' context.

A trigger for an axiom $\psi \equiv \forall \vec x. \phi(\vec x)$ is a
term (or a set of terms) that usually appears in $\phi(\vec x)$ and
which contains all the (term) variables $\vec x$ and all the type
variables in $\phi(\vec x)$. We use the notation
\[
\forall \vec x\;[\;p \;|\;
p_1, p_2].\; \phi(\vec x)
\]
to indicate that $\psi$ is associated with one mono-trigger $\{p\}$, and one
multi-trigger $\{p_1, p_2\}$. Triggers can either be provided by the user
with the syntax above, or heuristically computed by \aeg{}. In the
latter case, \aeg{} will choose at most two triggers per axiom by
default. For instance, possible triggers for the axiom {\tt mem\_add}
of Figure~\ref{fig:ex} are:

\begin{center}
  \{{\tt mem(x, add(y, s))} \} \qquad
  \{{\tt add(y, s), add(x, s)}\} \qquad
  $\cdots$ \qquad
  \{{\tt x, y, s}\}
\end{center}

The latest multi-trigger is a very bad choice and is never selected by
\aeg{}. In fact, it would generate an instance of {\tt mem\_add} for
every (well-typed) combination of terms appearing in the decision
procedures (\textit{resp.} SAT's model). The two first triggers seem
to be good choices. However, only the first one will permit us to
prove the validity of the example in Figure~\ref{fig:ex}. Indeed, the
ground term {\tt mem(a - 1, s2)} \textit{matches} the trigger {\tt
  mem(x, add(y, s))} modulo the equality {\tt s2 = add (b, s1)}. The
E-matching process produces the substitution $\{\mathtt{x \mapsto a -
  1,\; y \mapsto b,\; s \mapsto s1,\; \alpha \mapsto int}\}$, which
allows us to generate the needed instance.

\bigskip

The rest of the paper describes the features (and their implementation) that
\agr offers and shows how they can be useful both from an end-user perspective
as well as from a developer's perspective.

\section{Feedback}
\label{sec:feedback}

The first purpose of \agr is to provide \emph{feedback} which can be useful at
times to understand and evaluate what is happening inside the solver. Feedback
is useful for users as a visual aid to make sense of the solver's progress, but
it is also a precious tools for developers to profile and debug the solver.

\subsection{Unsat Cores and Minimal Context Extraction}
\label{sec:unsat}

An \emph{unsatisfiable core} in SMT, is a subset of the input formulas that
make the problem unsatisfiable. Traditionally, SMT solvers will return sets
where the elements are some of the input, top-level formulas, identified by a
unique name in the source. \aeg goes a bit further and identifies sub-formulas
that arise from the CNF (conjunctive normal form) conversion. This allows to
identify more precisely which part of the formula is actually useful in proving
the goal.

Unsat cores production is deactivated by default when running \aeg, but the
interface offers a way to change solver options on the fly, even while the
solver is running. In text, mode \aeg will spit unsat cores as pretty-printed
formulas on its output. This can become large at times. The interface will
display unsat cores in a more user-friendly way, visually identifying useful
parts of the context, hypotheses, \textit{etc.} by highlighting them in green
(see~Figure~\ref{fig:overview} for instance).

Different shades of green are used to highlight unsat cores in the buffer
window. Top-most declarations and definitions which contain part of the unsat
core will be highlighted in the lightest green and (sub-) formulas that appear
more frequently in the unsat core will be highlighted with a darker shade. In
particular, if an axiom is instantiated several times and the same part of the
resulting instances is actually useful to prove the goal, then the user will be
able to see this information visually. These features make it easy to rapidly
identify which parts of the context are useful and how crucial they are to
prove the goal.

Unsat cores also serve another purpose. By identifying which part of the
context is effectively used to prove the goal by the solver, we can remove any
other information contained in the problem while still having the guarantee
that the goal will be provable by the solver\footnote{This guarantee might be
  lost in some particular cases in \aeg, namely when an instance of a
  useless quantified axiom is used as a \emph{source of terms} to trigger the
  instantiation of another useful axiom.}. \agr offers a button in the toolbar
to quickly remove every top-level declaration or definition that does not
participate in the unsat core. Coupled with the mechanism of sessions (see
Section~\ref{sec:sessions}), this allows to save and replay already proven
goals much more rapidly.

\subsection{Instantiation}
\label{sec:instances}
\begin{figure}[htb]
  \centering
  \color{black!30}
  \frame{\includegraphics[width=.38\textwidth]{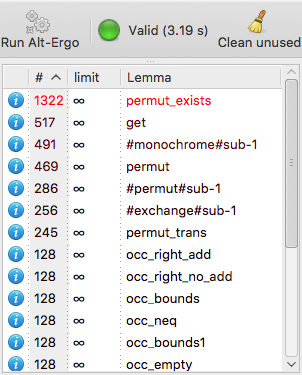}}
  \qquad
  \frame{\includegraphics[width=.38\textwidth]{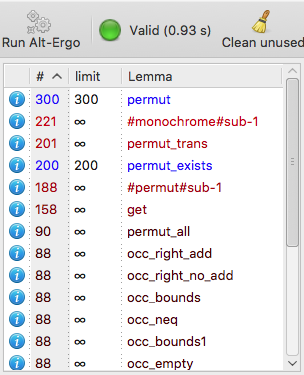}}
  % \vspace{-2em}
  \caption{Instances and manual limits}
  \label{fig:inst}
\end{figure}

As remarked in the introduction, quantified formulas are a source of
incompleteness and inefficiencies in most SMT solvers. Providing a way to
accurately and concisely expose information about instantiation is important
for the user experience. \agr does so by displaying axioms instantiated, in
real time, in a sub-window of the interface as shown in Figure~\ref{fig:inst}.

Here we report the number of instances produced by each axiom. They are listed
in decreasing order of number of instances and their names\footnote{For
  lemmas that are nested in larger formulas, we report the top-level name with
  some indication of their position. However, users can access their
  corresponding location in the source code by simply double clicking on the
  displayed name.}  are colored in varying shades of red to denote frequency of
instantiation. Axioms whose name is of a more saturated red denote the ones
which produce more instances with respect to the total number of instances
(regardless of its origin) generated at this point in time by the solver. This
allows to quickly identify potentially problematic axioms which generate too
many ground instances. This feedback gives indication regarding the likely
cause of problems in the instantiation mechanism.

When this happens, we also offer the possibility to \emph{limit} instantiation
of particular quantified lemmas. For example, the left report of
Figure~\ref{fig:inst} tells us that the lemma \textsf{permut\_exists} was
instantiated 1322 times, more than twice as much as any other lemma. This is
thus a good candidate to limit instantiation. On the right screen capture of
Figure~\ref{fig:inst}, we limited the instances of this problematic lemma to
200 and an associated lemma (\textsf{permut}) to 300 (we performed this process
iteratively, by first limiting instances of \textsf{permut\_exists} and looking
at what other lemmas were problematic). Lemmas for which instantiation has
reached its given limit are shown in blue. We can notice that the runtime of
the solver is reduced subsequently by a factor 3 for this particular example.

\subsection{Profiling}
\label{sec:profiling}

Much like in the spirit of the previous section, the bottom-right most
sub-window of the interface (see Figure~\ref{fig:overview}) gives \emph{real
  time} profiling information for the different modules and theories of
\aeg. These include the time spent in the SAT solver, the matching procedure,
the congruence closure algorithm (\textsf{CC(X)}), the builtin support for
associative and commutative symbols (\textsf{AC(X)})~\cite{acx} and the
theories of arithmetic, arrays, enumerated data-types (\textsf{Sum}) and
records.

\begin{figure}[htb]
  \centering
  \color{black!30}
  \frame{\includegraphics[width=.38\textwidth]{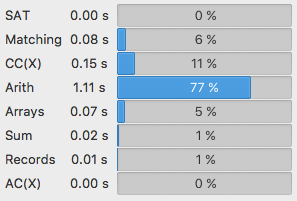}}
  \caption{Real time profiling information}
  \label{fig:prof}
\end{figure}

Figure~\ref{fig:prof} shows the state of the profiling information after
running an example. From the information displayed here, we can see that the
problem mostly stresses the theory of arithmetic in the solver. While the
solver is running, users can see which part takes the most time and can follow
the evolution. For instance, if the time reported for the theory of arrays is
too large in proportion and keeps growing, this can indicate that maybe there
are some axioms about arrays which are too permissive. Another use case, for
developers, is the possibility to identify problems where the solver is stuck
in a particular decision procedure (\textit{e.g.}, if only the timer for this
theory increases).

\section{Syntactic Operations}
\label{sec:syn}

A lot of the time when trying to use SMT solvers on real world examples (coming
from program verification tools for instance), the size of the logical context
and the sometimes heavy axiomatizations (that make liberal use of quantifiers
at varying degree of alternation) make the problem hard for purely automated
tools. However, only a fraction of the actual context is usually necessary to
prove the desired goal. Identifying useful information in very large problems
can be challenging. When the information provided by the feedback features
described in Section~\ref{sec:feedback} allow to identify a potential issue,
\agr offers a number of functionalities to perform syntactic manipulations on
the context that is shown in textual format. This allows for an iterative (and
slightly interactive) approach to SMT solving, where users can experiment and
quantify the effect of different manipulations.

\subsection{Pruning}
\label{sec:pruning}

Pruning can be performed very easily in \agr by simply double-clicking on the
top-level declaration or sub-formula that one wants to deactivate. Reactivating
previously deactivated items is also possible by performing the same action.

Most of the time pruning is not dangerous from a soundness point of
view. Removing an axiom only under-constrains the original problem which means
that any goal proved valid without some top-level hypothesis is still valid
with in the original context. However we also allow users to prune parts of a
formula. In this case validity can be affected depending on the
polarity of the
removed formula.

Consider the original goal:
\begin{tcb}{altergo}
logic P, Q : int -> prop
goal g: forall x, y: int. Q(x) or P(y) -> Q(x) and P(y)
\end{tcb}
which is trivially invalid. Removing \texttt{P(y)} on the left side of the
implication or on the right side of the implication changes the validity of
this goal. In fact removing both turns this goal into a valid one. \agr will
allow users to perform these potentially unsafe operations but will notify the
user by showing unsound prunings in red. A session that contains unsound
pruning operations cannot be saved either. This feature is still useful from an
end-user point of view because it allows to attempt proving goals by
strengthening hypotheses or weakening the goal itself. For instance if the goal
is a conjunction, a user can try to prove only part of the conjunction and
gather information from this attempt to help prove the rest.

\subsection{Dependency Analysis}
\label{sec:dep}

\agr maintains dependency information between declarations, definitions and
their use. It is possible to remove the declaration of a logical symbol and all
top-level declarations that make use of it in a single action. Conversely,
reactivating a previously pruned formula or declaration that uses a symbol will
also reactivate its declaration and/or definition.

A usage scenario for this feature, is to quickly disable a symbol for which we
know the axiomatization is problematic for the solver, then reactivate part of
the axiomatization iteratively in the hope that the solver will not be
overwhelmed anymore.

\subsection{Manual Instances}
\label{sec:manual}

Quantifiers are notoriously difficult for most SMT solvers. Unfortunately some
application domains such as deductive program verification make heavy use of
this feature to encode some domain specific functions. For instance
\textsf{Frama-C} has built-in axiomatizations for various memory models of
C. These are usually very large and complex.

Quantifier instantiation is a heuristic process for SMT solvers in general (although there
exists complete techniques for decidable fragments). \aeg uses \emph{matching
  modulo equality}. On the other side of the spectrum, interactive theorem
provers like \textsf{Coq} or \textsf{Isabelle} require users to perform
instantiation (\textit{i.e.} application) entirely manually. This is because in
traditional backward reasoning done in theorem provers, only relatively few applications
are necessary and a human can figure out which one to do based on the goal, the
context and knowledge about the current proof attempt.

\agr gives users the possibility to perform some instances manually. This is
useful for example if one has knowledge that a particular goal cannot be
solved without using specific instances of a lemma. \agr will ask for terms to
use in the instantiation. These can be constants but also other terms from the
context. Instances can also be \emph{partial}, which means that we allow that
only some of the variables be instantiated. Instances (partial or not) are
finally added to the context as hypotheses (they are shown in the top most
right corner of the window). All of the other presented actions can be
performed on instances.

\subsection{Triggers Selection and Modification}
\label{sec:triggers}

As mentioned in section~\ref{sec:matching}, \aeg computes
triggers---\textit{i.e.} patterns or filters used by the matching algorithm to
instantiate axioms---in a \emph{heuristic} way. For certain restricted
categories of axiomatizations, there exists techniques for computing triggers
that make the instantiation mechanism complete~\cite{dross16jar}, however this
is not the case in general and coming up with good triggers is a difficult
problem.

Because they essentially control how instances are generated, triggers play an
important role in proving goals with quantifiers. For example, consider the
following axiom, where \texttt{f} is an uninterpreted function from integers to
integers.
\begin{tcb}{altergo}
axiom idempotent : forall x : int. f(f(x)) = f(x)  
\end{tcb}
The trigger \texttt{f(f(x))} is more restrictive than \texttt{f(x)}. This means
that only terms $t$ that appear in larger terms \texttt{f(f(}$t$\texttt{))} will
be used for instantiating this axiom. Having this trigger will thus make the
solver generate \emph{less} instances of this axiom as opposed to the other one
possibility, \texttt{f(x)}.

There is a balance to be found when coming up with triggers, between restrictive
patterns and liberal ones. In some cases, the solver can be overwhelmed by the
instances generated if the triggers are not good, it will likely not terminate
from a user's point of view. If patterns are too restrictive, or if \aeg cannot
compute appropriate triggers, the affected axiom will likely not be
instantiated enough, preventing the solver of discovering potential
inconsistencies. In this latter case, \aeg will answer ``I don't know'' which
is usually unsatisfactory from an end user perspective.

Triggers can be specified by hand in the source problem for \aeg. The interface
\agr goes a step further by allowing triggers to be modified on the fly in the
source buffer window. The problem displayed to the user will show the triggers
computed by \aeg itself (it will also warn the user if the heuristic could not
come up with appropriate triggers for a quantified formula) as annotations in
the source. These can be modified interactively when one figures that the
heuristic did not produce the expected results.

\begin{figure}[htb]
  \centering
  \color{black!30}
  \frame{\includegraphics[width=.8\textwidth]{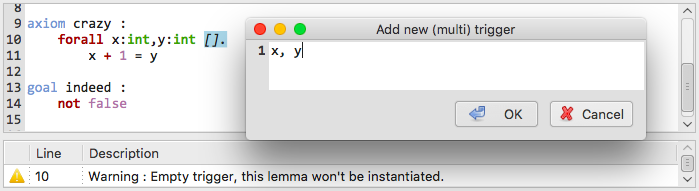}}
  % \vspace{-2em}
  \caption{Adding new triggers by hand}
  \label{fig:trigagr}
\end{figure}

Consider now the following, somewhat degenerate, goal.
\begin{tcb}{altergo}
 axiom crazy : forall x, y : int. x + 1 = y
 goal indeed: false 
\end{tcb}
\aeg will not compute any triggers for the axiom. By default, the solver
rejects triggers composed of single variables as they are considered too
permissive (any term of the appropriate type can be used for
instantiation). However in this specific case, we need to instantiate the axiom
with any two integers to discover the inconsistency. Figure~\ref{fig:trigagr}
shows\footnote{Goals in the interface are shown in their negated form, as they
  will be seen by the solver.} the functionality, right clicking on the trigger
displays a contextual menu to add triggers in a pop-up window. They can be
entered by the user, and \agr will then parse and type check the piece of text
corresponding to the trigger using the same internal functions (exposed) as the
ones of \aeg. In this example we add the multi-trigger \texttt{[x, y]} which
allows to conclude.

Triggers can also be deactivated (\textit{i.e.} removed from the axiom) by
using the same deactivation mechanism as the one for formulas. With these
possibilities, triggers can be modified at will without resorting on complex
solver parameters nor relying on heuristics.

\subsection{Sessions}
\label{sec:sessions}

Because this interface is designed to provide a slight increase in
interactivity for the user, we want all operations to be saved in what we call
\emph{sessions} for later \emph{replay}. In particular, operations that
manipulate the problem and custom tuning can greatly help the solver in its
search. When a user is happy with the state of its modifications---for
instance when they allowed to successfully prove the goal---the interface
offers the possibility to save the information concerning the session on disk.

This mechanism allows to replay sessions \emph{even if the original problem was
  modified}. Apart from the list of actions, the only additional information
that is saved in the session information is an association table between symbol
names for declarations and their identifiers (\texttt{id}). Node identifiers in
the AST are sequential integers following a depth-first ordering. When a
session file is read from the disk, the interface computes offsets for
identifiers using this association table in order to figure out the correct
corresponding ones for each action in the stack. Of course this is possible
only if modifications of the file are relatively minor. For instance, this will
work if some axioms were removed or added, or if a modification was performed
locally in a declaration (\textit{e.g.}, a formula was changed inside an
axiom). Nevertheless, sessions will not survive complete refactorings. If the
modifications are too important, the replay will try its best following this
offset strategy but can decide to abort if too many actions cannot be replayed.

Another usage of the session replay mechanism is to reuse sessions between
problems that share a lot of context. This is useful in a scenario where a user
found a satisfying set of modifications and tuning operations on a given
problem, and wants to try the same operations on a similar problem
(\textit{e.g.}, one where only the goal is different but the context is
identical).

\section{Implementation}

\begin{figure}[htb]
  \centering
  \includegraphics[width=0.8\textwidth]{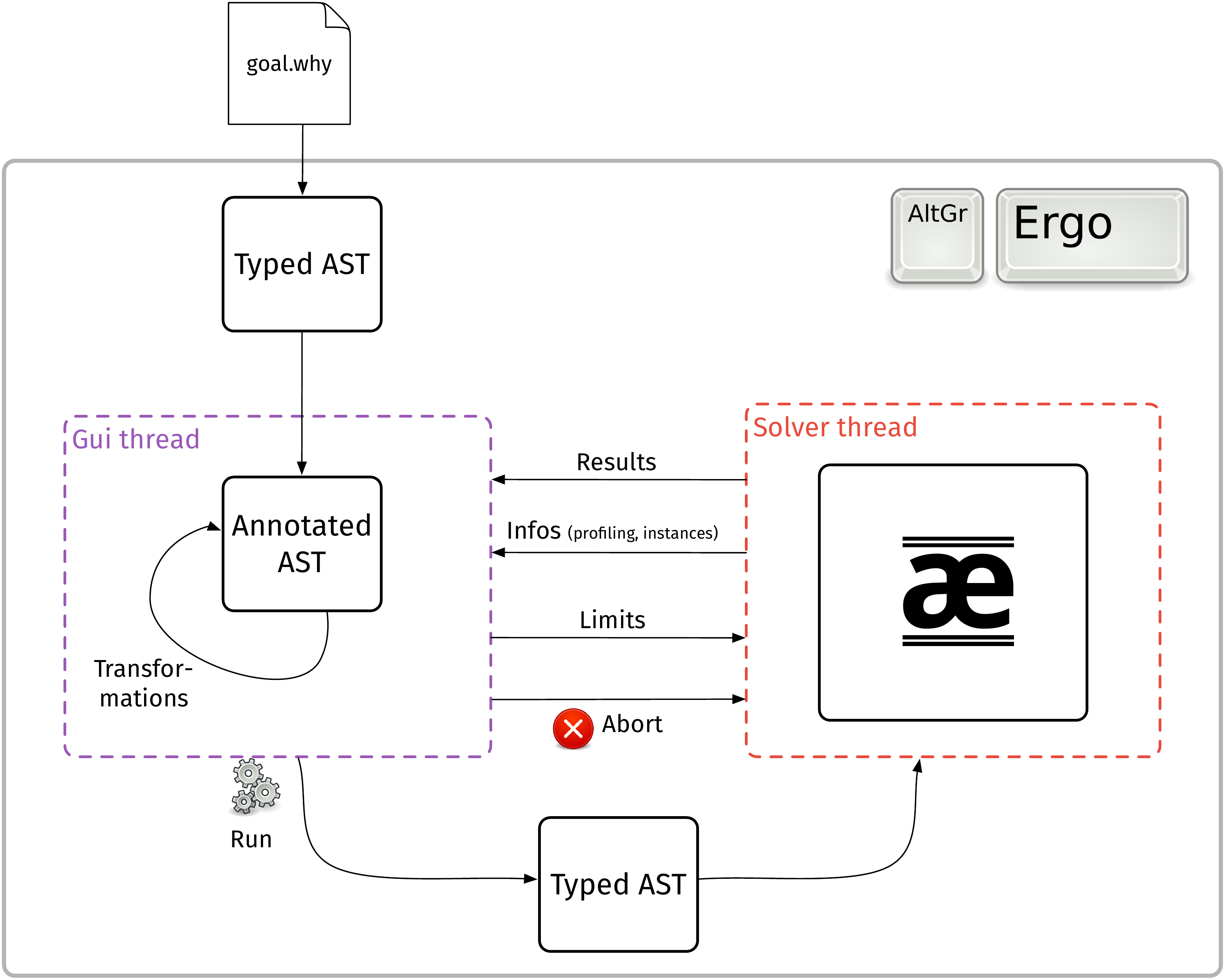}
  \caption{\agr's architecture}
  \label{fig:agrarchi}
\end{figure}

\agr is designed as a new front end for the solver. As such it reuses part of
\aeg's front end and \aeg's API. The interface is placed at the level of the
typed abstract syntax tree (Typed AST on Figure~\ref{fig:agrarchi}) and can
manipulate this representation at will. The solver itself communicates
various pieces of information to the interface.

The graphical part of the interface is written in GTK, using the OCaml bindings
LablGTK and runs in a separate thread. When \aeg is started in graphical mode,
one thread is created for the interface, and one thread is created for every
instance of the solver (started with the button ``Run''). These threads
communicate asynchronously through shared variables, messages or signals
depending on the functionality. For example, runs of the solver can be aborted
at anytime by clicking the button ``Abort'', a signal is emitted and caught by
the interface and the solver instance.

Most of the work performed by the interface is done on an \emph{annotated}
version of the AST.

\subsection{Annotations}

An annotated AST node is encoded by the record shown below. It contains the
node itself, mutable flags for pruning and polarities, GTK tags, a unique
identifier, the GTK buffer in which it is displayed and the line number at
which it appears in the source.
\begin{tcb}{ocaml}
type 'a annoted = {
  mutable c : 'a;          (* annotated value *)
  mutable pruned : bool;   (* pruned mark *)
  mutable polarity : bool; (* polarity for sub-terms *)
  tag : GText.tag;         (* GTK tag associated *)
  ptag : GText.tag;        (* Another GTK tag for proofs *) 
  id : int;                (* Unique identifier *)
  buf : sbuffer;           (* Source buffer *)
  mutable line : int;      (* line numbers can change *)
}
\end{tcb}
Some of these fields are mutable to account for user actions. For instance
pruning a sub-term in the source amounts to changing the flag
\texttt{pruned}. In turn, polarities can be affected by these operations (see
Section~\ref{sec:pruning}).

Tags are used as a way to effectively identify portions of the source code
displayed to the user in the buffer. These special tags of type
\lstinline[language=ocaml]{GText.tag}, allow to efficiently change properties
of the display at any moment. For instance they are used to show the user
sub-terms of the formulas and their type when the mouse is hovered over the
particular location in the buffer area. Tags can be stacked, this is why we
have a second one, \texttt{ptag}, which displays unsat cores and remains
unaffected by any other action (see Section~\ref{sec:unsat}). The operations
described in Section~\ref{sec:syn} are all performed on this intermediate
representation.

When the user clicks on the button \emph{Run}, annotations are stripped form
the AST and the resulting AST is sent to a newly created thread for this
specific instance of the solver \aeg.

\subsection{Sessions}

Every action performed in the interface is saved in a stack in the current
environment. This is the stack that is exported when the session is saved.
\begin{tcb}{ocaml}
type action =
  | Prune of int
  | IncorrectPrune of int
  | Unprune of int
  | AddInstance of int * string * string list
  | AddTrigger of int * bool * string
  | LimitLemma of int * string * int
  | UnlimitLemma of int * string

type env =
  { ...
    saved_actions : action Stack.t;
    ... }
\end{tcb}

Actions are parameterized by the identifier of the AST node on which they were
performed. Only atomic actions are saved. This is because all other operations
can be expressed in terms of these atomic actions.  This simple structure
allows to replay sessions at a later date and obtain a state which is
equivalent to the one that was shown to the user when it was saved on disk.

\section{Conclusion and Future Work}

We've presented \agr, a graphical front end for the SMT solver \aeg. It
provides real time feedback to users and allows to interactively manipulate
problems. We believe such a tool is beneficial both from an end-user and
developer point of view, especially for the kind of goals that arise from
deductive program verification.  We have personally (as the developers of \aeg)
found the interface precious to tune the solver and tackle larger verification
problems.

Future directions to improve this tool include turning \agr into a full
featured editor for SMT solvers (at the moment the source buffer window cannot
be edited directly). There are still some inefficiencies in the interface when
one wants to load \emph{very large} files, due to some of the more advanced
features, but also inherent to GTK itself. Another future direction for this
tool, would be to dissociate the interface from the prover completely with the
objective of providing a \emph{web based interface}, which would be more
portable, flexible and easy to use. Also, the session mechanism could be
further improved to allow for more replay scenarios by integrating some more
advanced diff and merge techniques~\cite{autexier}.

This work is a first step towards \emph{semi-interactive} theorem provers. One
more ambitious goal would be to provide a set of \emph{tactics} and an
appropriate language to allow users to do part of their proof manually, or to
finish (also partly manually) proofs that the SMT solver struggle with. This
would involve work both at the level of the proof language as well as the level
of the user interface.

\bibliographystyle{eptcs}
\bibliography{biblio}

\begin{thebibliography}{10}
\providecommand{\bibitemdeclare}[2]{}
\providecommand{\surnamestart}{}
\providecommand{\surnameend}{}
\providecommand{\urlprefix}{Available at }
\providecommand{\url}[1]{\texttt{#1}}
\providecommand{\href}[2]{\texttt{#2}}
\providecommand{\urlalt}[2]{\href{#1}{#2}}
\providecommand{\doi}[1]{doi:\urlalt{http://dx.doi.org/#1}{#1}}
\providecommand{\bibinfo}[2]{#2}

\bibitemdeclare{article}{autexier}
\bibitem{autexier}
\bibinfo{author}{Serge \surnamestart Autexier\surnameend}
  (\bibinfo{year}{2015}): \emph{\bibinfo{title}{Similarity-Based Diff,
  Three-Way Diff and Merge}}.
\newblock {\sl \bibinfo{journal}{International Journal of Software \&
  Informatics}} \bibinfo{volume}{9}(\bibinfo{number}{2}).
\newblock
  \urlprefix\url{http://www.ijsi.org/ch/reader/view_abstract.aspx?file_no=i217}.

\bibitemdeclare{book}{spark}
\bibitem{spark}
\bibinfo{author}{John \surnamestart Barnes\surnameend} (\bibinfo{year}{2012}):
  \emph{\bibinfo{title}{{SPARK}: The Proven Approach to High Integrity
  Software}}.
\newblock \bibinfo{publisher}{Altran Praxis},
  \bibinfo{address}{http://www.altran.co.uk, UK}.

\bibitemdeclare{inproceedings}{AE_polymorphism}
\bibitem{AE_polymorphism}
\bibinfo{author}{Fran\c{c}ois \surnamestart Bobot\surnameend},
  \bibinfo{author}{Sylvain \surnamestart Conchon\surnameend},
  \bibinfo{author}{Evelyne \surnamestart Contejean\surnameend} \&
  \bibinfo{author}{St\'ephane \surnamestart Lescuyer\surnameend}
  (\bibinfo{year}{2008}): \emph{\bibinfo{title}{{Implementing Polymorphism in
  SMT solvers}}}.
\newblock In: {\sl \bibinfo{booktitle}{SMT '08/BPR '08: Proceedings of the
  Joint Workshops of the 6th International Workshop on Satisfiability Modulo
  Theories and 1st International Workshop on Bit-Precise Reasoning}},
  \bibinfo{publisher}{ACM}, \bibinfo{address}{New York, NY, USA}, pp.
  \bibinfo{pages}{1--5}, \doi{10.1145/1512464.1512466}.
\newblock \urlprefix\url{http://www.lri.fr/~conchon/publis/conchon-smt08.pdf}.

\bibitemdeclare{manual}{AtelierB}
\bibitem{AtelierB}
\bibinfo{organization}{ClearSy System Engineering}:
  \emph{\bibinfo{title}{Atelier B User Manual, version~4.0}}.
\newblock
  \urlprefix\url{http://tools.clearsy.com/wp-content/uploads/sites/8/resources/User_uk.pdf}.

\bibitemdeclare{article}{acx}
\bibitem{acx}
\bibinfo{author}{Sylvain \surnamestart Conchon\surnameend},
  \bibinfo{author}{Evelyne \surnamestart Contejean\surnameend} \&
  \bibinfo{author}{Mohamed \surnamestart Iguernelala\surnameend}
  (\bibinfo{year}{2012}): \emph{\bibinfo{title}{Canonized Rewriting and Ground
  AC Completion Modulo Shostak Theories : Design and Implementation}}.
\newblock {\sl \bibinfo{journal}{Logical Methods in Computer Science}}
  \bibinfo{volume}{8}(\bibinfo{number}{3}), \doi{10.2168/LMCS-8(3:16)2012}.
\newblock
  \urlprefix\url{http://www.lri.fr/~conchon/publis/conchon-lmcs2012.pdf}.

\bibitemdeclare{article}{ccx}
\bibitem{ccx}
\bibinfo{author}{Sylvain \surnamestart Conchon\surnameend},
  \bibinfo{author}{Evelyne \surnamestart Contejean\surnameend},
  \bibinfo{author}{Johannes \surnamestart Kanig\surnameend} \&
  \bibinfo{author}{St\'ephane \surnamestart Lescuyer\surnameend}
  (\bibinfo{year}{2008}): \emph{\bibinfo{title}{CC(X): Semantic Combination of
  Congruence Closure with Solvable Theories}}.
\newblock {\sl \bibinfo{journal}{Electronic Notes in Theoretical Computer
  Science}} \bibinfo{volume}{198}(\bibinfo{number}{2}), pp.
  \bibinfo{pages}{51--69}, \doi{10.1016/j.entcs.2008.04.080}.
\newblock
  \urlprefix\url{http://www.lri.fr/~conchon/publis/conchon-entcs08.pdf}.

\bibitemdeclare{inproceedings}{AE-RSSR-2016}
\bibitem{AE-RSSR-2016}
\bibinfo{author}{Sylvain \surnamestart Conchon\surnameend} \&
  \bibinfo{author}{Mohamed \surnamestart Iguernelala\surnameend}
  (\bibinfo{year}{2016}): \emph{\bibinfo{title}{Increasing Proofs Automation
  Rate of Atelier-B Thanks to Alt-Ergo}}.
\newblock In \bibinfo{editor}{Thierry \surnamestart Lecomte\surnameend},
  \bibinfo{editor}{Ralf \surnamestart Pinger\surnameend} \&
  \bibinfo{editor}{Alexander \surnamestart Romanovsky\surnameend}, editors:
  {\sl \bibinfo{booktitle}{Reliability, Safety, and Security of Railway
  Systems. Modelling, Analysis, Verification, and Certification - First
  International Conference, RSSRail 2016, Paris, France, June 28-30, 2016,
  Proceedings}}, {\sl \bibinfo{series}{Lecture Notes in Computer Science}}
  \bibinfo{volume}{9707}, \bibinfo{publisher}{Springer}, pp.
  \bibinfo{pages}{243--253}, \doi{10.1007/978-3-319-33951-1}.

\bibitemdeclare{article}{dross16jar}
\bibitem{dross16jar}
\bibinfo{author}{Claire \surnamestart Dross\surnameend},
  \bibinfo{author}{Sylvain \surnamestart Conchon\surnameend},
  \bibinfo{author}{Johannes \surnamestart Kanig\surnameend} \&
  \bibinfo{author}{Andrei \surnamestart Paskevich\surnameend}
  (\bibinfo{year}{2016}): \emph{\bibinfo{title}{Adding Decision Procedures to
  {SMT} Solvers using Axioms with Triggers}}.
\newblock {\sl \bibinfo{journal}{Journal of Automated Reasoning}}
  \bibinfo{volume}{56}(\bibinfo{number}{4}), pp. \bibinfo{pages}{387--457},
  \doi{10.1007/s10817-015-9352-2}.

\bibitemdeclare{inproceedings}{why3}
\bibitem{why3}
\bibinfo{author}{Jean-Christophe \surnamestart Filli{\^a}tre\surnameend} \&
  \bibinfo{author}{Andrei \surnamestart Paskevich\surnameend}
  (\bibinfo{year}{2013}): \emph{\bibinfo{title}{{Why3 - Where Programs Meet
  Provers}}}.
\newblock In: {\sl \bibinfo{booktitle}{ESOP}}, pp. \bibinfo{pages}{125--128},
  \doi{10.1007/978-3-642-37036-6\_8}.

\bibitemdeclare{phdthesis}{thesisMohamed}
\bibitem{thesisMohamed}
\bibinfo{author}{Mohamed \surnamestart Iguernelala\surnameend}
  (\bibinfo{year}{2013}): \emph{\bibinfo{title}{Strengthening the heart of an
  SMT-solver : Design and implementation of efficient decision procedures.
  (Renforcement du noyau d'un d{\'{e}}monstrateur {SMT} : Conception et
  implantation de proc{\'{e}}dures de d{\'{e}}cisions efficaces)}}.
\newblock Ph.D. thesis, \bibinfo{school}{University of Paris-Sud, Orsay,
  France}.
\newblock \urlprefix\url{https://tel.archives-ouvertes.fr/tel-00842555}.

\bibitemdeclare{article}{framac2015}
\bibitem{framac2015}
\bibinfo{author}{Florent \surnamestart Kirchner\surnameend},
  \bibinfo{author}{Nikolai \surnamestart Kosmatov\surnameend},
  \bibinfo{author}{Virgile \surnamestart Prevosto\surnameend},
  \bibinfo{author}{Julien \surnamestart Signoles\surnameend} \&
  \bibinfo{author}{Boris \surnamestart Yakobowski\surnameend}
  (\bibinfo{year}{2015}): \emph{\bibinfo{title}{Frama-C: A software analysis
  perspective}}.
\newblock {\sl \bibinfo{journal}{Formal Aspects of Computing}}
  \bibinfo{volume}{27}(\bibinfo{number}{3}), pp. \bibinfo{pages}{573--609},
  \doi{10.1007/s00165-014-0326-7}.

\bibitemdeclare{article}{e-matching}
\bibitem{e-matching}
\bibinfo{author}{Micha{\l} \surnamestart Moskal\surnameend},
  \bibinfo{author}{Jakub \surnamestart {\L}opusza\'{n}ski\surnameend} \&
  \bibinfo{author}{Joseph~R. \surnamestart Kiniry\surnameend}
  (\bibinfo{year}{2008}): \emph{\bibinfo{title}{{E}-matching for Fun and
  Profit}}.
\newblock {\sl \bibinfo{journal}{Electron. Notes Theor. Comput. Sci.}}
  \bibinfo{volume}{198}, pp. \bibinfo{pages}{19--35},
  \doi{10.1016/j.entcs.2008.04.078}.
\newblock \urlprefix\url{http://dl.acm.org/citation.cfm?id=1371256.1371287}.

\end{thebibliography}

\end{document}